\documentclass{wscpaperproc}
\usepackage{latexsym}
\usepackage{graphicx}
\usepackage{mathptmx}
\usepackage[T1]{fontenc}

\usepackage{amsmath}
\usepackage{amsfonts}
\usepackage{amssymb}
\usepackage{amsbsy}
\usepackage{amsthm}

\usepackage{xcolor}
\definecolor{LightGray}{gray}{0.95}
\usepackage{minted}
\usepackage{algorithm}
\usepackage{algorithmicx}
\usepackage{algpseudocode}
\usepackage[utf8]{inputenc} % allow utf-8 input
\usepackage{url}            % simple URL typesetting
\usepackage{booktabs}       % professional-quality tables
\usepackage{nicefrac}       % compact symbols for 1/2, etc.
\usepackage{listings}
\newcommand{\xb}{\mathbf{x}}
\newcommand{\Xb}{\mathbf{X}}
\newcommand{\yobs}{\mathbf{y}_{\text{obs}}}
\usepackage[pdftex,colorlinks=true,urlcolor=blue,citecolor=black,anchorcolor=black,linkcolor=black]{hyperref}

\newtheoremstyle{wsc}% hnamei
{3pt}% hSpace abovei
{3pt}% hSpace belowi
{}% hBody fonti
{}% hIndent amounti1
{\bf}% hTheorem head fontbf
{}% hPunctuation after theorem headi
{.5em}% hSpace after theorem headi2
{}% hTheorem head spec (can be left empty, meaning `normal')i

\theoremstyle{wsc}

\begin{document}

%***************************************************************************
% AUTHOR: AUTHOR NAMES GO HERE
% FORMAT AUTHORS NAMES Like: Author1, Author2 and Author3 (last names)
%
%		You need to change the author listing below!
%               Please list ALL authors using last name only, separate by a comma except
%               for the last author, separate with "and"
%

% setting up general page style
\pagestyle{fancyplain}

% setting up page style of first page
\thispagestyle{plain}
\firstPageHead{}

% setting up running header (authors) of subsequent pages
\chead{\fancyplain{}{\itshape O'Gara, Fadikar, Binois, Collier, and Ozik}}

% setting up seperation parameters
%\headsep=72pt
\rhead{}
\cfoot{}
\renewcommand{\headrulewidth}{0pt} % (renewcommand needed in fancyhdr to remove top decorative line)
%\headrulewidth=0pt  % ("setlength" needed in fancyheading to remove top decorative line)

%%%%%%%%%%%%%%%%%%%%%%%%%%%%%%%%%%%%%%%%%%%%%%%%%%%%%%%%%%%%%%%%%%%%%%%%%%%%%%
%                                                                            %
%     THESE COMMANDS ARE REQUIRED TO WORK WITH WSC.BST TO MAKE BIBLIO     %
%                                                                            %
%%%%%%%%%%%%%%%%%%%%%%%%%%%%%%%%%%%%%%%%%%%%%%%%%%%%%%%%%%%%%%%%%%%%%%%%%%%%%%
\makeatletter
\let\@internalcite\cite
\def\cite{\def\@citeseppen{-1000}%
    \def\@cite##1##2{(##1\if@tempswa , ##2\fi)}%
    \def\citeauthoryear##1##2##3{##1 ##3}\@internalcite}
\def\citeNP{\def\@citeseppen{-1000}%
    \def\@cite##1##2{##1\if@tempswa , ##2\fi}%
    \def\citeauthoryear##1##2##3{##1 ##3}\@internalcite}
\def\citeN{\def\@citeseppen{-1000}%
%  Pierre L'Ecuyer's fix for multiple cite bug
%  Added by Paul J Sanchez on 4 October 2001
%   \def\@cite##1##2{##1\if@tempswa , ##2)\else{)}\fi}%
%   \def\citeauthoryear##1##2##3{##1 (##3}\@citedata}
    \def\@cite##1##2{##1\if@tempswa, ##2)\else{}\fi}%
    \def\citeauthoryear##1##2##3{##1 (##3)}\@citedata}
\def\citeA{\def\@citeseppen{-1000}%
    \def\@cite##1##2{(##1\if@tempswa , ##2\fi)}%
    \def\citeauthoryear##1##2##3{##1}\@internalcite}
\def\citeANP{\def\@citeseppen{-1000}%
    \def\@cite##1##2{##1\if@tempswa , ##2\fi}%
    \def\citeauthoryear##1##2##3{##1}\@internalcite}
\def\shortcite{\def\@citeseppen{-1000}%
    \def\@cite##1##2{(##1\if@tempswa , ##2\fi)}%
    \def\citeauthoryear##1##2##3{##2 ##3}\@internalcite}
\def\shortciteNP{\def\@citeseppen{-1000}%
    \def\@cite##1##2{##1\if@tempswa , ##2\fi}%
    \def\citeauthoryear##1##2##3{##2 ##3}\@internalcite}
\def\shortciteN{\def\@citeseppen{-1000}%
%  Pierre L'Ecuyer's fix for multiple cite bug
%  Added by Paul J Sanchez on 2 September 2002
%  should have caught this last year...
%   \def\@cite##1##2{##1\if@tempswa , ##2)\else{)}\fi}%
%   \def\citeauthoryear##1##2##3{##2 (##3}\@citedata}
% Shane G. Henderson fix for extra right bracket at end of optional material June 8, 2005
%    \def\@cite##1##2{##1\if@tempswa, ##2)\else{}\fi}%
    \def\@cite##1##2{##1\if@tempswa, ##2\else{}\fi}%
    \def\citeauthoryear##1##2##3{##2 (##3)}\@citedata}
\def\shortciteA{\def\@citeseppen{-1000}%
    \def\@cite##1##2{(##1\if@tempswa , ##2\fi)}%
    \def\citeauthoryear##1##2##3{##2}\@internalcite}
\def\shortciteANP{\def\@citeseppen{-1000}%
    \def\@cite##1##2{##1\if@tempswa , ##2\fi}%
    \def\citeauthoryear##1##2##3{##2}\@internalcite}
\def\citeyear{\def\@citeseppen{-1000}%
    \def\@cite##1##2{(##1\if@tempswa , ##2\fi)}%
    \def\citeauthoryear##1##2##3{##3}\@citedata}
\def\citeyearNP{\def\@citeseppen{-1000}%
    \def\@cite##1##2{##1\if@tempswa , ##2\fi}%
    \def\citeauthoryear##1##2##3{##3}\@citedata}
%
% \@citedata and \@citedatax:
%
% Place commas in-between citations in the same \citeyear, \citeyearNP,
% \citeN, or \shortciteN command.
% Use something like \citeN{ref1,ref2,ref3} and \citeN{ref4} for a list.
%
\def\@citedata{%
    \@ifnextchar [{\@tempswatrue\@citedatax}%
                  {\@tempswafalse\@citedatax[]}%
}

\def\@citedatax[#1]#2{%
\if@filesw\immediate\write\@auxout{\string\citation{#2}}\fi%
  \def\@citea{}\@cite{\@for\@citeb:=#2\do%
    {\@citea\def\@citea{, }\@ifundefined% by Young
       {b@\@citeb}{{\bf ?}%
       \@warning{Citation `\@citeb' on page \thepage \space undefined}}%
{\csname b@\@citeb\endcsname}}}{#1}}%

% don't box citations, separate with ; and a space
% also, make the penalty between citations negative: a good place to break.
%
\def\@citex[#1]#2{%
\if@filesw\immediate\write\@auxout{\string\citation{#2}}\fi%
  \def\@citea{}\@cite{\@for\@citeb:=#2\do%
    {\@citea\def\@citea{; }\@ifundefined% by Young
       {b@\@citeb}{{\bf ?}%
       \@warning{Citation `\@citeb' on page \thepage \space undefined}}%
{\csname b@\@citeb\endcsname}}}{#1}}%

% (from apalike.sty)
% No labels in the bibliography.
%
\def\@biblabel#1{}
\makeatother

%\newlength{\bibhang}
%\setlength{\bibhang}{2em}

% Indent second and subsequent lines of bibliographic entries. Taken
% from openbib.sty: \newblock is set to {}.
% \renewcommand{\refname}{REFERENCES}

\newdimen\bibindent
\bibindent=0.0em
% SEC: was \def\thebibliography#1{\section*{\refname\@mkboth
% SEC: was   {\uppercase{\refname}}{\uppercase{\refname}}}\list
\def\thebibliography#1{\section*{\refname}\list
   {}{\settowidth\labelwidth{[#1]}
   \leftmargin\parindent
   \itemindent -\parindent
   \listparindent \itemindent
   \itemsep 0pt
   \parsep 0pt}
   \def\newblock{}
   \sloppy
   \sfcode`\.=1000\relax}

           % Set up BiBTeX macros

% needed to make the tex document look more like the word counterpart :-(
\setlength{\baselineskip}{12.7pt}

% AUTHOR: Enter the title, all letters in upper case
\title{Trajectory-Oriented Optimization Via Adaptive Thompson Sampling And Grid Refinement: A Tutorial With The ADAPTIVE\_TS Package}

% AUTHOR: Enter the authors of the article, see end of the example document for further examples
\author{\begin{center}David O'Gara\textsuperscript{1}, Arindam Fadikar\textsuperscript{1}, Micka\"{e}l Binois\textsuperscript{2}, Nicholson Collier\textsuperscript{1}, and Jonathan Ozik\textsuperscript{1}\\
[11pt]
\textsuperscript{1}Decision and Infrastructure Sciences, Argonne National Laboratory, Lemont IL, USA\\
\textsuperscript{2}Inria, Université Côte d’Azur, CNRS, LJAD, Sophia Antipolis, FRANCE\\
\end{center}
}

\maketitle

\vspace{-12pt}

\section{ABSTRACT}
Stochastic simulators are increasingly used to expand the frontier of scientific knowledge and inform decision-making across real-world contexts. Simulator calibration, a process by which internal model inputs are tuned to match some external criteria, usually in the form of observed data, is a key step in model design and validation. Epidemiological simulators present an especially compelling use case, as evidenced by the recent COVID-19 pandemic. Among several calibration paradigms, trajectory-oriented optimization is an emerging approach that does not require assumptions on the stochastic behavior of the simulator replicates and is particularly effective at identifying trajectories through the lens of errors between the simulator and observed data, especially when combined with Bayesian optimization. We present a tutorial on trajectory-oriented optimization with \texttt{adaptive\_ts}, an open-source Python package. We also provide a series of worked examples on an accompanying webpage.

\section{Introduction}

Calibrating stochastic simulations to empirical data remains an ongoing challenge for mathematical modelers and decision-makers across myriad scientific fields~\shortcite{baker2022analyzing,reiker2021emulator,fadikar2023trajectory,fadikar2025adaptive,ozik2021population,gramacy2020surrogates,kennedy2001bayesian}. Of particular interest is the calibration of epidemiological simulators, for at least three reasons: (1) the usage of simulation models for crisis management and outbreak response, from avian influenza~\shortcite{germann2006mitigation,eubank2004modelling}, to COVID-19 more recently~\shortcite{ozik2021population,adam2020special}; (2) infectious disease dynamics represent open, nonlinear, noisy systems, making them a challenge to model and calibrate accurately~\shortcite{anderson1991infectious,vespignani2020modelling,ionides2006inference}; (3) the increased availability and democratization of high-performance computing resources and simulation software, making it possible to utilize these types of models to answer detailed research questions~\shortcite{collier2024distributed,collier2022distributed,ozik2025automation,binois2025portfolio}. However, the methods and tools used to calibrate these models are often developed for a particular use case or individual simulator, which is partly due to epidemic models being developed under high-stakes scenarios and short timelines to inform decision-making~\shortcite{ozik2021population,ferguson2020report}. Further, competing paradigms exist on the precise definition of calibration for a stochastic epidemiological model~\shortcite{DOG2025improving}. One common approach is to identify parameters that, when simulated many times, recover the data in expectation~\shortcite{baker2022analyzing}. An alternative perspective is to instead focus on acquiring individual simulation \textit{trajectories} via coupled parameter and random seed pairings~\shortcite{fadikar2023trajectory} without making distributional assumptions about the underlying parameters. This approach has not yet been widely studied partly due to methodological and implementation challenges requiring the use of common random numbers, a variance-reduction approach in numerical methods, that has not typically been supported by default by most modeling approaches~\shortcite{chen2012effects,fadikar2023trajectory}. The trajectory-based approach is particularly well-suited to path-dependent epidemic simulators such as agent-based models (ABMs)\shortcite{fadikar2025adaptive,kerr2021controlling}. The latter perspective, which we term ``Trajectory Oriented Optimization (TOO)'' is the focus of this work~\shortcite{fadikar2023trajectory,pearce2022bayesian}. Here, we present a tutorial of TOO as applied to stochastic epidemiological simulation models and as operationalized via the Adaptive Thompson Sampling (\texttt{adaptive\_ts}) Python package. The methods, problems, and data demonstrated in this paper are available as a public repository and as an \href{https://github.com/emews/wsc2026-TOO-tutorial}{accompanying website}~\shortcite{WSC_2026_TOO_2026}.

Our priorities when designing the \texttt{adaptive\_ts} package, as well as developing this tutorial were the following:
\begin{itemize}
    \item Flexibility: beyond several packages from the standard scientific Python stack, we prioritize having few software dependencies. In particular, we recognize that designing detailed stochastic simulators requires significant software development in their own right, or are run on HPC resources where individual users have less control over their dependencies, and thus for maximal utility, any additional tools should come with minimal installation requirements.
    \item HPC workflow integration: HPC workflow software, such as EMEWS, decouples the model exploration algorithms from the simulator and enables large-scale simulation-based analyses~\shortcite{collier2024distributed,ozik2018extreme}.
    \item Customization: recognizing that individual use cases may arise depending on the calibration problem at hand, we allow users the option to write their own methods for fitting surrogate models, in a similar modular approach that has become popular in recent years for machine learning libraries like \texttt{PyTorch}~\shortcite{paszke2019pytorch}, \texttt{BoTorch}~\shortcite{balandat2020botorch}, and \texttt{GPJax}~\shortcite{pinder2022gpjax}.
    
\end{itemize}

The remainder of this paper is organized as follows. In the following section, we review related work to TOO and discuss a motivating example using a pedagogical epidemiological simulation. We provide an overview of each of the required components for a workflow within the \texttt{adaptive\_ts} package along with examples. We then perform a detailed case study using a real simulation model developed to inform decision-making for the Chicago Department of Public Health (CDPH). Finally, we conclude  with a discussion of limitations, challenges, and opportunities for future work.

\section{Related Work and Motivating Example}

Various methodological approaches exist to calibrate stochastic epidemiological simulators to data, such as Approximate Bayesian computation (ABC)~\shortcite{sisson2007sequential}, Markov Chain Monte Carlo (MCMC)~\shortcite{gelman1996markov} and Simulation-Based Inference (SBI)~\shortcite{radev2020bayesflow}. The unifying paradigm among these inference approaches is that they, as statistical models, rely on the assumption of long-run behaviors under many simulations: given enough simulations, a well-specified inference process should be able to identify which model parameterizations \textit{in expectation} are more consistent with the data. While this view is important, calibration becomes more difficult when the simulator at hand exhibits high variance in the outcome or path-dependence within a simulation. Stated more simply, if repeated runs of the same parameter value $x$ in a simulator resulted in widely different outbreak sizes (e.g., $y_i(x) = 100$ and $y_j(x) = 1,000$), then calibrating a simulator and choosing parameters based on the ``average'' outbreak size may have limited utility. This is in part due to a loss of identifiability between the simulation parameters and simulation outcomes and, in more practical terms, in the potentially limited ability to inform stakeholder decision-making.

Instead, recent work on ``Trajectory-Oriented Optimization'' (TOO)~\shortcite{fadikar2023trajectory} posits simulator stochasticity as something to be explicitly learned and utilized instead of a nuisance to be averaged over. TOO accomplishes this by adjusting the learning goal from learning about model parameter values $\xb$ to instead discovering pairs of parameter and random seed values $(\xb,r)$ that induce the desired trajectories that are similar to empirical data. For many classes of epidemiological simulators, the simulation process can be expensive, in some cases even requiring multiple hours for a single simulation~\shortcite{aylett2021june}, meaning that extensive experimentation with the simulator is not always feasible or the most appropriate use of available resources. This is especially the case where the time-to-solution to meaningfully inform decision-making is at a premium~\shortcite{ozik2025automation}. Thus, TOO relies on the use of surrogate or \textit{emulator} models: statistical approximations that can make uncertainty-aware predictions on yet-to-be simulated parameter values~\shortcite{gramacy2020surrogates}, along with Bayesian Optimization (BO)~\shortcite{garnett2023bayesian} to acquire promising combinations of parameters and random seeds in a sample-efficient manner. BO also requires an acquisition function that determines which model simulations to acquire subject to an exploration of potentially new parameter space locations versus exploitation of known acquisitions. Our preferred method, Thompson sampling (TS)~\shortcite{thompson1933likelihood}, elegantly balances exploration and exploitation by sampling from the posterior predictive distribution and thus does not require additional tuning parameters. TS also natively supports potentially large batches of parameter evaluations, facilitating the use of concurrency on HPC resources. Finally, TS also has the auxiliary benefit of not requiring  a continuous domain for optimization, making it particularly suitable for a mixed-domain problem such as TOO, containing continuous simulator parameters and a discrete set of random seeds.

\begin{figure}[ht!]
    \centering    \includegraphics[width=0.95\linewidth]{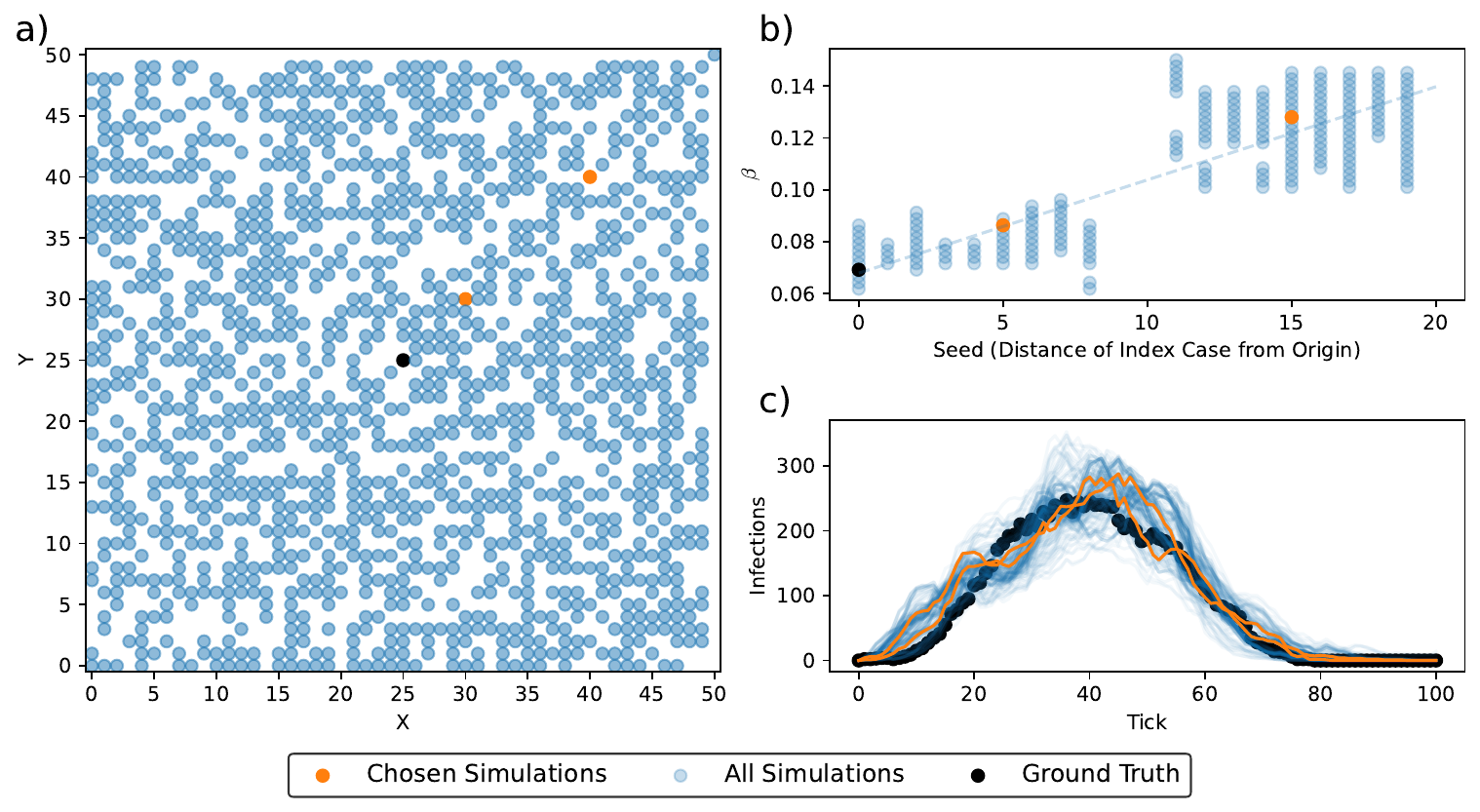}
    \caption{Identifiability tradeoffs in a simple SIR model: (a) the starting model configuration for all simulation runs: the black dot denotes the ground truth of an index case starting in the center, and the two orange dots show alternate locations of the index case (specifically 30,30 and 40,40  denoting random seeds of 5 and 15). (b) explicitly shows the tradeoff between distance from the center and corresponding transmission probabilities of the closest-matching simulations (for example, index cases farther from the center have fewer transmission opportunities early in the model, and thus require larger transmission probabilities to yield trajectories like the ground truth). (c) shows the actual trajectories and demonstrates their alignment with the data.}
    \label{fig:repastSIR}
\end{figure}

To motivate the utility of TOO and the interplay between model parameters and random seeds, consider the following simplified example from computational epidemiology: we instantiate a population of $N = 2000$ agents on a two-dimensional grid with coordinates $(x,y)$ between 0 and 50, and create a Susceptible-Infected-Recovered (SIR) model of disease dynamics. We designate all the agents to be Susceptible and randomly choose one of them to be the index Infected~\shortcite{anderson1991infectious}. Each agent takes a random walk on the grid at each time step, and Infected agents may probabilistically, with per-contact transmission probability $\beta$, pass their infection to nearby Susceptible ones, who then become Infected for a fixed number of time steps. Thus, stochasticity in this version of the SIR model comes from three places: (1) the mixing pattern induced by the random walks among the agents, (2) the probabilistic infection process, and (3) the placement of the index infection. To demonstrate the role of the random seed and its utility for TOO methods, we isolate the stochasticity associated with the placement of the index case. This is motivated by the common pattern in ABM epidemiological modeling where the placement of initial infected individuals in a population of agents is often governed by a random seed to capture the uncertainty in infection importation processes, i.e., where and when new infections are introduced into a population. We then use an identical random number stream to govern the rest of the model stochasticity across all the experiments, namely the agent random walks and the infection processes, minimizing the across-simulation stochasticity effects outside of the index case location assignment. We use the following convention for placement of the index case: a random seed of 0 places the index case at the center of the grid $(25,25)$, and each increment of the random seed places the index case along the right diagonal of the grid, so a random seed of $r$ corresponds to a placement of $(25+r, 25 + r)$ on the grid. A priori, we expect that placing the index case closer to the center should lead to larger outbreaks for the same $\beta$ value since the index agent in the center of the grid can most easily reach any subsequent point across the grid via a series of random walks. Our implementation is in \texttt{Repast4Py}~\shortcite{collier2022distributed,collier2020experiences} and hereafter we refer to this model as \texttt{repastSIR}. We perform an experiment with \texttt{repastSIR} where we simulate over a dense grid of $\beta$ values and 20 random seeds and record the difference between a chosen ground truth ($\beta^* \approx 0.069, r^* = 0$) and the simulated data. We keep simulations with a root mean-squared-error (RMSE) below 40 to denote trajectories that are sufficiently close to the data for our purposes, and note that a specific RMSE cutoff will depend on the application. Figure~\ref{fig:repastSIR} contains the results of this experiment, showing a clear identifiability tradeoff between $\beta$ values and placement of the index case in panel (b). We plot these collected trajectories in panel (c) to show their similarity to the ground truth data. However, absent additional information, it is challenging to know what to \textit{do} with these trajectories to inform decision-making. Further, since we fixed the randomness associated with agent movement across all of these simulations, it is difficult to know if these same $\beta$ parameters values will produce trajectories similar to the ground truth for different movement patterns. In this work we posit that it would be of interest for modelers to be able to better quantify the effect of stochasticity in their computational models by incorporating both the ``regular'' model parameters as well as the random seeds.

\section{Adaptive\_TS Workflow}

The Adaptive TS workflow requires three components: (1) the emulator, which is trained on simulation data and can make uncertainty-aware predictions at yet-to-be simulated parameterizations, (2) a grid refinement strategy that alongside the emulator assess the suitability of candidate acquisitions and performs the sampling, and (3) a seed expansion strategy that determines when to expand the discrete, but countably infinite, random seed space. These are combined into a \texttt{Workflow} class and carry out the adaptive sampling procedure, where promising parameter and seed combinations are repeatedly acquired. Throughout, we assume the following: (1) that the continuous simulation parameters lie on a bounded unit hypercube $\mathcal{X} \in [0,1]^d$, which follows standard best practice in emulator modeling~\shortcite{gramacy2020surrogates,balandat2020botorch} and facilitates space-filling designs such as a Latin Hypercube~\shortcite{mckay2000comparison}. We also assume (2) that the random seed space may without loss of generality be characterized by a set of integers $Z \in [1,..,k]$ and that the seeds don't have an ordinal relationship. The current version of the Adaptive TS workflow is best suited for scalar simulator outputs. In the case of the epidemiological model examples shown in this work, we typically compute the sum of squared errors $\sum_{i=1}^N(Y_i - Y_{obs_i})^2$ between the time-series simulator output $Y$ and observed data $Y_{obs}$ and thus aim to minimize this discrepancy, as has been done in other work~\shortcite{fadikar2023trajectory,fadikar2025adaptive,fadikar2025developing}. Typically, we also apply a log transform and standardize the output for ease of use with Gaussian process emulators. The package also supports some in-development methods for multi-output emulators, which is a useful approach for modeling high-dimensional simulation outputs, such as time-series~\shortcite{fadikar2018calibrating,higdon2008computer,robertson2025bayesian}.

\subsection{Emulation}

The emulator represents the approximation function for the stochastic simulator. In the \texttt{adaptive\_ts} package, the emulator requires methods for fitting to input and output data (\texttt{fit}), making predictions at potential input locations (\texttt{predict}), and sampling from the posterior distribution (\texttt{sample}). Several candidate emulators are provided with the package, in particular the ones from the \texttt{hetGPy} package~\shortcite{DOG2025hetgpy}, including the common random number Gaussian process model \texttt{crnGP}, a Gaussian process regression model introduced in \cite{fadikar2023trajectory} that is specifically made to model continuous parameters along with random seeds, which is available in \texttt{adaptive\_ts} as \texttt{crnGPEmulator}. In Figure~\ref{fig:emulator-class} we show the code for the \texttt{BaseEmulator} class. 

Figure~\ref{fig:gp-emulator} demonstrates the basic steps for how one would implement a specific emulator from another Python package, using the popular \texttt{GaussianProcessRegressor} from \texttt{scikit-learn} as an example use case. We note that a full implementation of the \texttt{scikit-learn} GPR is available with the package as \texttt{GaussianProcessEmulator}. We also support several other emulators with the \texttt{adaptive\_ts} package as extensions, which currently include \texttt{TabPFN}, an advanced neural network capable of state-of-the art predictions on tabular data~\shortcite{hollmann2025accurate}, as well as our implementation of the intrinsic coregionalization model (ICM)~\shortcite{bonilla2007multi}, which is available as \texttt{SeedlingGP}. The details of our \texttt{SeedlingGP} implementation are discussed in the case study in Section \ref{casestudy}.

\begin{figure}[!h]
    %\centering
    \begin{minipage}[frame=single,fontsize=\tiny]{0.8\textwidth}
\small
\begin{verbatim}
import numpy as np

class BaseEmulator:
    """
    Base class to create an emulator.
    Requires implementing a fit, 
    predict, and sample method
    """
    def __init__(self,rng: np.random.Generator = None):
        self.rng = rng
    def __str__(self):
        return "BaseEmulator"
    def fit(self,X: np.ndarray,Y: np.ndarray):
        raise NotImplementedError()
    def predict(self) -> dict:
        raise NotImplementedError()
    def sample(self) -> np.ndarray:
        raise NotImplementedError()
\end{verbatim}

\end{minipage}
    \caption{The \texttt{BaseEmulator} class showing the required \texttt{fit}, \texttt{predict}, and \texttt{sample} methods, which are inspired by the \texttt{scikit-learn} API.}
    \label{fig:emulator-class}
\end{figure}

\begin{figure}[!h]
    %\centering
    \begin{minipage}[frame=single]{0.8\textwidth}
\small
\begin{verbatim}
import numpy as np
from sklearn.gaussian_process import GaussianProcessRegressor as GPR

class GaussianProcessEmulator(BaseEmulator):
    """
    Wrapper around scikit-learn GaussianProcessRegressor (GPR)
    """
    def __init__(self, kernel: kernels):
        super().__init__()
        self.kernel = kernel
        self.model = GPR(kernel=kernel)
    def __str__(self):
        return "sklearnGP"
    def fit(self, X: np.ndarray, Y: np.ndarray):
        self.model.fit(X, Y)
    def predict(self, X: np.ndarray, return_cov: bool):
        return self.model.predict(X, return_cov=return_cov)
    def sample(self, X: np.ndarray, size:int =1):
        mean, cov = self.predict(X=X, return_cov=True)
        return self.rng.multivariate_normal(mean, cov, size)
\end{verbatim}

\end{minipage}

%\caption{A sample implementation of an emulator with the \texttt{scikit-learn} \texttt{GaussianProcessRegressor} model, implementing the \texttt{fit}, \texttt{predict}, and \texttt{sample} methods.}
    \caption{A sample implementation of an emulator with the \texttt{scikit-learn}\newline\texttt{GaussianProcessRegressor} model, implementing the \texttt{fit}, \texttt{predict}, and \texttt{sample} methods.}
    \label{fig:gp-emulator}
\end{figure}

\subsection{Grid Refinement}

Given an emulator, the next component of an Adaptive TS workflow is the grid from which candidate locations are proposed for the next acquisition, a process that we refer to as the ``grid strategy''. Each grid strategy must implement a \texttt{sample} method. Further, the grid strategy also interacts with the emulator via its \texttt{predict} and \texttt{sample} methods. Each grid strategy also requires the following parameters:

\begin{itemize}
    \item \texttt{ndim} (integer): the number of continuous model parameters
    \item \texttt{nseeds} (integer): the number of seeds that comprise the discrete search space
    \item \texttt{ngrid} (integer): the grid size (default 100)
\end{itemize}

In the following, we describe the types of grid strategies available in the package. We note that since all of the grid strategies assume the continuous parameters lie on a unit hypercube and that the seeds are described by integers, users are required to keep track of two components for their simulators: (1) re-scaling of the inputs to their native units (which may easily be done via, e.g., \texttt{scipy.stats.qmc.rescale()}) and (2) converting seeds from the seed search space to simulator seeds (which may be easily accomplished with a lookup table or mapping such as a \texttt{dict} in Python).

\subsubsection{Non-Adaptive Grids}

The \texttt{grid} module supports two types of non-adaptive grids, the \texttt{FixedGrid} and \texttt{LHSGrid}. The \texttt{FixedGrid} does not allow for sampling of new candidate parameter locations across the Adaptive TS workflow acquisitions. Users have the option to use a one-time Latin Hypercube sample or to provide an explicit grid of candidate locations upon instantiation. The fixed grid formulation is useful when simulations utilize physical phenomena, such as specific locations, as in~\cite{sun2019}, where simulated and field data were combined to model solar irradiance. The \texttt{LHSGrid} strategy allows for sampling from a new Latin Hypercube design at each iteration, which offers additional exploration beyond a \texttt{FixedGrid}. However, the \texttt{LHSGrid} strategy still only samples from the whole unit hypercube.  It does not utilize information from the prior acquisitions to focus on regions of interest, thus motivating the use of adaptive grid strategies, which we discuss next.

\subsubsection{Adaptive Grids}

In many BO or simulator calibration workflows, the parameter search space may be characterized by a broad range of behaviors, and the parameter ranges consistent with desired outcomes may occupy a small region of the total parameter volume~\shortcite{ozik2021population,kerr2021controlling,andrianakis2015bayesian,vernon2010galaxy}. This motivates the use of adaptive grid strategies, which promote sampling near the region of interest, thus making it more likely that the grid strategy will acquire promising candidate parameterizations. The \texttt{adaptive\_ts} package offers two built-in adaptive grid strategies, the \texttt{AdaptiveGridStrategy}, which implements the approach from~\shortcite{fadikar2025adaptive} and the \texttt{NormalizingFlowsStrategy}. The \texttt{AdaptiveGridStrategy} operates in two stages, which we refer to as filtering and densification. First, the grid strategy draws a Latin Hypercube sample, and then the filtering step performs an importance sampling using the emulator to remove unlikely parameterizations. The densification step involves a series of Metropolis-Hastings-like proposals $\xb_{\text{can}} \sim q(\cdot \mid \xb)$ that allow for perturbations around the filtered grid while also enabling exploration.

This procedure is summarized as pseudocode below (reproduced from~\shortcite{fadikar2025adaptive}):

\begin{algorithm}[!ht]
\caption{Pseudo-code for Adaptive-grid generation with fixed seed set $\mathcal{R}$}\label{algo:grid}
\begin{algorithmic}[1]
\Require  $\tilde{\Xb}^{(0)}_t = \{(\xb_1, r_1), \dots, (\xb_M, r_M)\}$ (grid of size $M$), fixed seed set $\mathcal{R} = \{r_1, \dots, r_l\}$

\For{$i = 1, \dots, M$}
    \State $w(\xb_i, r_i) \gets L\big((\xb_i, r_i) \mid \yobs, \mathcal{D}_t, \cdot\big)$
    \State $w'(\xb_i, r_i) \gets w(\xb_i, r_i) / \sum_{j=1}^M w(\xb_j, r_j)$
\EndFor
\State Resample $(\xb_i, r_i)$ with probability $w'(\xb_i, r_i)$ (with replacement) to obtain $\tilde{\Xb}^F_{t}$
\State Initialize $\tilde{\Xb}_{t}^{\text{adapt}} \gets \tilde{\Xb}^F_{t}$
\State $m \gets |\tilde{\Xb}_{t}^{\text{adapt}}|$
\While{$m < M$}
    \State Choose $(\xb, r) \in \tilde{\Xb}_{t}^{\text{adapt}}$ at random
    \State Sample $\xb_{\text{can}} \sim q(\cdot \mid \xb)$
    \For{$r_j \in \mathcal{R}$}
        \State Compute $\alpha = \min\left\{1, \frac{L\big((\xb_{\text{can}}, r_j) \mid \yobs, \mathcal{D}_t, \cdot\big) \cdot q(\xb_{\text{can}} \mid \xb)}{L\big((\xb, r) \mid \yobs, \mathcal{D}_t, \cdot\big) \cdot q(\xb \mid \xb_{\text{can}})}\right\}$
        \State Sample $u \sim \text{Uniform}(0, 1)$
        \If{$u < \alpha$ and $(\xb_{\text{can}}, r_j) \notin \tilde{\Xb}_{t}^{\text{adapt}}$}
            \State $\tilde{\Xb}_{t}^{\text{adapt}} \gets \tilde{\Xb}_{t}^{\text{adapt}} \cup (\xb_{\text{can}}, r_j)$
            \State $m \gets m + 1$
            \State \textbf{break}
        \EndIf
    \EndFor
\EndWhile
\State Return $\tilde{\Xb}_{t}^{\text{adapt}}$ as the updated search grid
\end{algorithmic}
\end{algorithm}

The \texttt{NormalizingFlowStrategy} also utilizes an importance sample to filter the candidate acquisition points, and then samples directly from an approximate posterior $\xb_{\text{can}} \sim q_F(\cdot | \xb)$ via a trained normalizing flow~\shortcite{papamakarios2021normalizing,nflows}. The default normalizing flow used is a masked autoregressive flow~\shortcite{papamakarios2017masked}. Since normalizing flows and, more broadly, deep learning approaches in general,~\shortcite{cranmer2020frontier}, typically require extensive training data, which may be at odds with the low-sample regime of Bayesian optimization~\shortcite{garnett2023bayesian}, we also offer the option to save the normalizing flow at each iteration so it may be retrained on subsequent data as opposed to fitting a new flow at every iteration.

\subsection{Random Seed Expansion}

An Adaptive TS workflow begins with a finite set of random seeds to pair with a model's parameters. Random seed expansion involves adding new potential random seed locations to the search space. In the Adaptive TS workflow, this involves adding a new seed to the existing set of random seeds. We offer the following heuristics as potential ways for a user to expand their seed space:
\begin{itemize}
    \item A user may want to expand their seed space after every $n$ simulations, implemented in the class \texttt{SeedExpansionBySims}
    \item A user may want to have a fixed probability $p$ of seed expansion at every iteration, implemented in the class \texttt{SeedExpansionByProb}
    \item A user may want to expand their seed space based on some user-defined criteria as a function of the emulator, the acquired data, and the current size of the seed space, and thus they would extend the provided \texttt{BaseSeedExpansion} class to implement this
\end{itemize}

The \texttt{BaseSeedExpansion} class also implements a basic \texttt{sample} method that generates a number of samples from the refined grid (supplied by the grid strategy) and replaces each sample's seed with the newest (largest) seed so it may be added to the emulator training data. This represents a pure exploration approach when expanding the seed space, but users are free to implement their own, arbitrary \texttt{sample} methods, such as changing the seed of the current ``best'' trajectories and swapping in the newest expanded seed, which would reflect a more exploitative (rather than exploratory) approach. We intentionally leave these implementations up to the user because we anticipate that the choice of seed expansion depends heavily on the simulator at hand, since the choice of a new seed may correspond simply to a different mixing pattern in an epidemiological simulation, or in some cases may involve a more intensive process, such as generating a new synthetic population in a more complex model~\shortcite{DOGTraceSTL2025}. In any implementation of seed expansion, the user is required to supply the following parameters to the seed expansion class:

\begin{itemize}
    \item \texttt{nseeds} (integer) the initial size of the seed space (if not supplied, it is taken from the grid strategy)
    \item \texttt{nexpansion} (integer) the number of samples to draw from the refined grid when calling the seed expansion \texttt{sample} method
\end{itemize}

As can be seen above, in designing the seed expansion classes, we expose both the emulator and grid strategy to the seed expansion class, offering users maximum flexibility in their design.

\subsection{Integrated Adaptive TS Workflow}

Each of the above components (the emulator, grid strategy, and seed expansion protocol) are combined into a \texttt{Workflow} class that handles the interconnections between these classes. The key inputs to the \texttt{Workflow}, other than the above component parts, are the simulator \texttt{budget} which tells the workflow how many total simulations to run, and a pseudo-random number generator \texttt{rng}, which is also passed to the emulator and grid strategies upon workflow instantiation, unless other random number generators were supplied to each of those classes. The \texttt{Workflow} class also contains the methods \texttt{fit\_surrogate} (which calls the \texttt{fit} method of the emulator module), \texttt{sample\_grid\_strategy} to call the \texttt{sample} from the grid strategy module, \texttt{check\_for\_expansion} to see if the seed expansion condition is satisfied, and \texttt{sample\_from\_expansion} to generate new samples from the expanded seed space. Figure~\ref{fig:sample-workflow}  demonstrates how all of the components fit together into a workflow.

\begin{figure}[ht!]
    %\centering
    %\begin{minted}[bgcolor=LightGray]{python}
    \begin{minipage}{0.8\textwidth}

    \small
    \begin{verbatim}
from scipy.stats.qmc import LatinHypercube
from adaptive_ts.emulator import crnGPEmulator
from adaptive_ts.grid import AdaptiveGridStrategy
from adaptive_ts.expansion import SeedExpansionBySims
from adaptive_ts.data import Dataset # container class to store data
from adaptive_ts.utils import make_grid # utility to merge parameters and seeds

ndim = 1 # 1 continuous parameter
lhs = LatinHypercube(d=ndim)
emulator = crnGPEmulator(covtype="Matern5_2")
adapt_strategy = AdaptiveGridStrategy(
            ndim = ndim, nseeds = 5,  # 1 continuous parameter, 5 seeds
            ngrid = 100, nTS_samp=30 # 100 pars to acquire, 30 Thompson samples
)
expansion_strategy = SeedExpansionBySims(
    nseeds = 5, # 5 random seeds to start
    nsims_expand = 50 # expand seed space every 50 simulations,
    nexpansion = 10 # number of new samples to draw when seed space is expanded
)
workflow = Workflow(
        emulator = emulator,
        grid_strategy = adapt_strategy,
        seed_expansion_strategy = expansion_strategy,
        budget = 200, # total simulation budget
        rng = np.random.default_rng(1) # random number generator
)
X0, Y0 = get_data("...") # read in some starting data
nsims = len(Y0)
data = Dataset(X=X0, Y=Y0)

while nsims < workflow.budget: # run workflow until budget is reached
    X, Y = data.X, data.Y
    workflow.fit_surrogate(X, Y)
    newX = workflow.sample_grid_strategy()
    if workflow.check_for_expansion():
        X_expanded_seeds = workflow.sample_from_expansion(newX)
        newX = np.concatenate([newX, X_expanded_seeds], axis=0)
    new_out = run_batch(newX, ytrue=ytrue) # run simulations
    newY = new_out["Y"]
    data.append(X=newX, Y=newY)
    nsims += len(newY)
    \end{verbatim}
    \end{minipage}
    \caption{Sample workflow, specifying the emulator (crnGP), grid strategy (adaptive), and seed expansion (expansion by sims) to use.}
    \label{fig:sample-workflow}
\end{figure}

\begin{figure}[!h]
    \centering
    \includegraphics[width=0.9\linewidth]{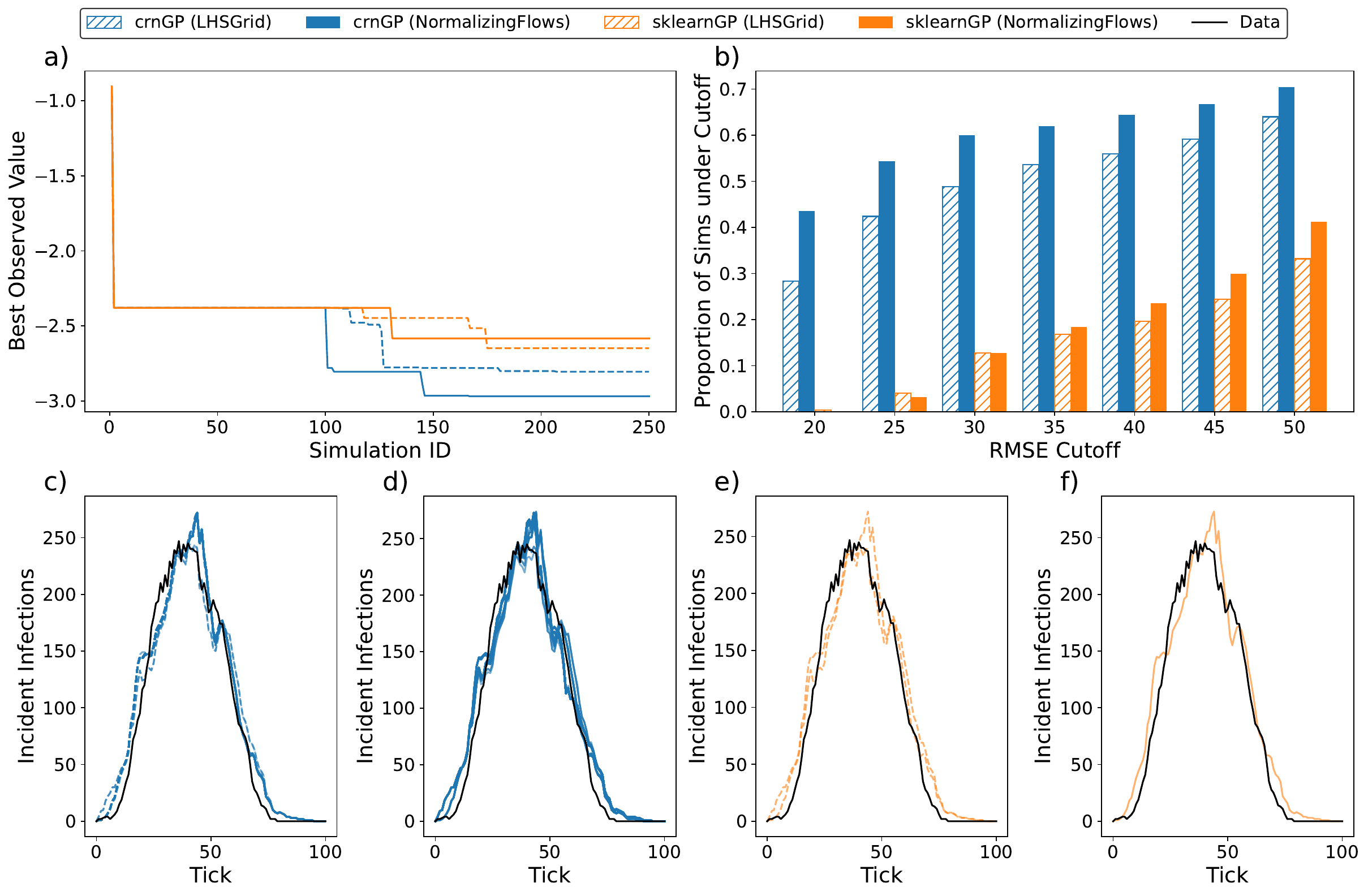}
    \caption{Summary of Adaptive TS workflow for repastSIR showing two emulators and two grid strategies: (a) the best observed value (error between ground truth and simulation output, lower is better) across simulation runs, stratified by experiment, (b) workflow efficiency, with a higher proportion of simulator runs below a cutoff indicating that the workflow is using its model runs efficiently, (c)-(f) the actual simulation trajectories which have an RMSE below 20. The scikit-learn based GPs are less efficient and identify far fewer eligible trajectories than the crnGP based ones, and the use of an adaptive grid (in this case, via normalizing flows) also identifies more suitable trajectories.}
    \label{fig:repastSIR-workflow}
\end{figure}

Figure~\ref{fig:repastSIR-workflow} shows a set of sample workflow results, varying across two emulators and two grid strategies, as applied to the \texttt{repastSIR} simulator introduced earlier. To recap, we simulate a ``ground truth'' with an infection probability $\beta \approx 0.069$ and an index location in the center of a 50x50 grid, with a random seed search space of $r \in [1,20]$, meaning that the ground truth $r^*$ is not contained in the search space. In analyzing the quality of a workflow, we are particularly interested in three results. The best observed value, which in this case is the log of the sum of squared errors (and then standardized), is shown in panel (a), which tracks the minimal value of the objective across simulated trajectories. We are also interested in generating a large number of high-quality trajectories (measured by RMSE between simulated and ground truth infections) as a proportion of our total simulation budget (in this case, 200 simulations), as shown in panel (b). Finally, we also check that the high-quality (low RMSE) trajectories qualitatively capture the key features of the ground truth simulation, namely having about the same peak infections at about the same time in the case of an epidemiological model, which are shown in panels (c)-(f).  Across the different workflows, the key observations are that the choice of emulator and grid strategy impact not only the overall optimization procedure (i.e., obtaining different best observed values) but also the sample efficiency and proportion of high-quality trajectories found under a limited simulation budget. We see that using a strategy tailored for TOO (the \texttt{crnGP} emulator) paired with an adapted grid refinement strategy (the \texttt{NormalizingFlows} class) outperforms less TOO inclined strategies such as a standard homoskedastic Gaussian process emulator and non-adaptive grid strategies such as Latin Hypercube sampling.

% \newpage

\section{Case Study}\label{casestudy}

In this section, we present a case study using the \texttt{MetaRVM} simulation model~\shortcite{fadikar2025developing,fadikar_metarvm_2025}, which was developed to inform decision-making for the Chicago Department of Public Health (CDPH). \texttt{MetaRVM} is a metapopulation model tailored for respiratory virus simulation, such as influenza, COVID-19, or RSV, and the version utilized here is stratified into six geographic areas representing the \href{https://www.chicago.gov/city/en/depts/cdph/supp_info/healthy-communities/healthy-chicago-zones.html}{``Healthy Chicago''} zones (HCEZs)~\cite{ChiHCEZ}. \texttt{MetaRVM} utilizes a set of mixing matrices based on the Chicago Social Interaction Model (ChiSIM)~\shortcite{macal2018chisim}, a framework that simulates a synthetic population of Chicago with every individual following an hourly schedule, which includes movement between key locations such as homes, workplaces, and schools. The mixing matrices model key population flows and interactions within and across the HCEZs, and thus, where the simulated epidemic begins can impact the geospatial progression and risk across each of the population strata. See We calibrate two model parameters, the transition rate from Susceptible to Exposed \texttt{ts} and the number of initially exposed individuals $E_{ini}$, which have parameter search ranges of $\text{ts} \in [0.05,0.5]$ and $E_{ini} \in [2500,25000]$ in a total population of nearly 2.56 million. We also model a third parameter, which is a categorical variable denoting the distribution of the initial infected agents among the six geographic zones. As emphasized throughout this tutorial, the TOO paradigm attempts to model meaningful randomness induced by the structure of the simulator. An outbreak starting in Chicago's North/Central zone, containing a population of 683,220 and commuter-heavy workplace locations such as the Loop, will likely have a different epidemic progression than one instantiated in the Northwest or Southwest zones, which have smaller populations, but contain O'Hare and Midway International Airports, respectively, and thus represent two additional importation pathways. In what we present next, we show how the TOO approach allows us to disambiguate between these importation modalities. By generating sets of trajectories that more closely align with importation-informed dynamics, the model-based analysis can better inform decision-making regarding additional surveillance or interventions across the zones. This would be much more challenging to address with a non-TOO approach, as any epidemic behaviors arising from the initial outbreak zone would likely either be unmodeled by the emulator (and thereby attributed to noise) or, worse, confounded with model parameters.   We generate 7 total initial infected distributions, six of which put the entire distribution in one zone, and one that distributes them uniformly across the zones, and are listed below:
\begin{table}[]
    \centering
    \begin{tabular}{lrrrrrr}
     \toprule
     Experiment & North/Central & Near South & Northwest & Southwest & West & Far South\\
     \midrule
     1 & $1$ & $0$ & $0$ & $0$ & $0$ & $0$ \\
      2 & $0$ & $1$ & $0$ & $0$ & $0$ & $0$ \\
       3 & $0$ & $0$ & $1$ & $0$ & $0$ & $0$ \\
        4 & $0$ & $0$ & $0$ & $1$ & $0$ & $0$ \\
         5 & $0$ & $0$ & $0$ & $0$ & $1$ & $0$ \\
          6 & $0$ & $0$ & $0$ & $0$ & $0$ & $1$ \\
 7 & $\nicefrac{1}{6} $ & $\nicefrac{1}{6} $ & $\nicefrac{1}{6} $ & $\nicefrac{1}{6} $ & $\nicefrac{1}{6} $ & $\nicefrac{1}{6}$ \\
 \bottomrule
 \end{tabular}

    \caption{Summary of initial infected distribution for each of six HCEZ locations in Chicago, Illinois. Experiments 1-6 denote having the entire outbreak beginning in one zone, while Experiment 7 distributes the number of initial infections uniformly across all zones.}
    \label{tab:metarvm-table}
\end{table}

Our key question is whether we can reliably recover the correct ground truth of which zone contains the initial infections in addition to trajectories. This would have implications for retrospective analyses in supporting decision-makers, in that recovering the geospatial structure could inform deployment of additional surveillance resources to mitigate epidemic spread. We generated each of the seven ground truth simulations across a range of parameters and total initial infections, and we utilized a seed search space containing seven distributions of initial infections and five different random number streams, for a total seed space of 35 random seeds. Since we imbued the seed space with more structure than just independent random number streams, we utilized the \texttt{SeedlingGP} emulator in the \texttt{adaptive\_ts} package, which is a variant of the (ICM)~\shortcite{bonilla2007multi}, which is made for multi-task modeling. Specifically, we re-purpose the implementation of the \texttt{MultiTaskGP} model available in \texttt{GPyTorch}~\shortcite{gardner2018gpytorch}, where each random seed would be a different (but correlated) output task. The \texttt{SeedlingGP} implementation instead handles this mapping by treating the random seed column as the input to the task correlation matrix, and thus the covariance between two inputs $\xb_1 = (x_1,r_1)$ and $\xb_2 = (x_1,r_2)$ is:
\begin{align*}
    k(\xb_1,\xb_2) = k_\text{continuous}(x_1,x_2)\times k_{\text{seed}}(r_1,r_2),
\end{align*}

where $k_{\text{continuous}}$ is any stationary kernel in \texttt{GPyTorch} such as the RBF or Mat\'ern (in our case study, we utilize the Mat\'ern kernel with $\nu = \nicefrac{5}{2}$), and $k_{\text{seed}}$ is a variant of the \texttt{IndexKernel} in \texttt{GPyTorch} which is of the form:
\begin{align*}
    k(i,j) = \left(BB^T + \text{diag}(\mathbf{v})\right)_{i,j},
\end{align*}
where $B$ is a low-rank matrix and $\mathbf{v}$ is non-negative, with the additional constraint that $BB^T$ has a unit diagonal. The seed kernel formulation allows for modeling of a high-dimensional input space, which is this case comprises 37 dimensions: two input parameters and 35 random seeds, which begins to push the traditional suggested boundaries for BO~\shortcite{garnett2023bayesian,frazier2018tutorial}. If the random seed was treated like a normal categorical variable, via one-hot encoding, this would make interpreting distances (and thus lengthscale hyperparameters) in such a high-dimensional space challenging. Additionally, a one-hot encoding treats the levels of a categorical variable as independent from one another, while in our formulation, we expect some correspondence between subsets of random seeds due to their common geographic importation structure.

\begin{figure}[ht!]
    \centering
    \includegraphics[width=0.8\textwidth]{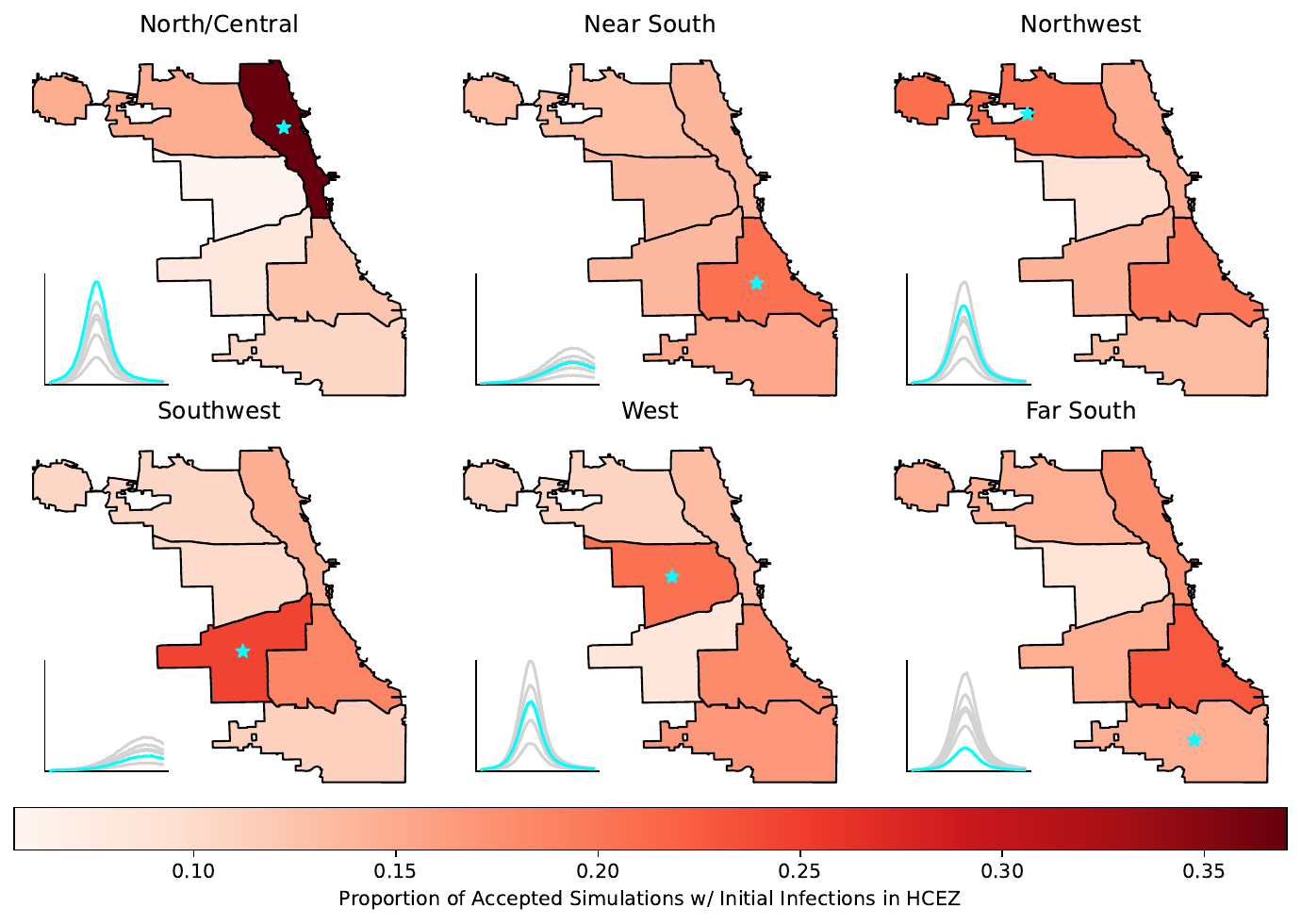}
    \caption{Summary of accepted trajectories by HCEZ for six ground truth experiments, shown in each panel. Each experiment comprised 700 total simulations, of which 350 were acquired after the initial design, and the acceptance criteria was an aggregate RMSE of 500 between trajectories and ground truth data for the acquired simulations. The ground truth trajectories are shown as inset plots in each panel, with the cyan trajectory corresponding to the HCEZ with the initial infections, and others shown in gray.}
    \label{fig:chicago-map}
\end{figure}

For each ground truth experiment, we design an initial set of 350 simulations, comprising 10 continuous parameterizations and the 35 seeds, and acquire 350 additional simulations with the AdaptiveTS workflow, using normalizing flows as the grid strategy. We set a cutoff of 500 for the RMSE between simulated and ground truth data for the acquired simulations, aggregated across the geographic zones to define a trajectory as ``accepted''. Figure~\ref{fig:chicago-map} shows the proportion of the accepted trajectories starting in each of the six HCEZs. The proportion of the accepted trajectories in each HCEZ may be thought of as the agreement between simulations beginning in that zone with the ground truth data: thus, a higher proportion of accepted trajectories in a given HCEZ would indicate a higher belief that the epidemic began in that location. The results demonstrate that we are able to detect  geographic patterns of infection importation, especially when the epidemic starts in the North/Central zone, and the Southwest. For the Far South ground truth experiment, we notice that the adjacent zone (Near South) has the most accepted trajectories. This is consistent with the underlying mixing matrices that show considerable mixing between these two zones~\shortcite{WSC_2026_TOO_2026}. In the case of the North/Central zone, instantiating the epidemic there has particularly high signal recovery since the North/Central zone contains Chicago's Loop, which is the central business district and thus has the highest mixing between zones.

\section{Conclusion}

In this work, we have demonstrated the utility for the Adaptive TS workflow and provided a tutorial using the \texttt{adaptive\_ts} package. Throughout, we have emphasized the contributions of the components of an Adaptive TS workflow (the emulator, grid strategy, and seed expansion strategy) as well as reflections on how to tailor each component to individual use cases. We have also provided a case study using a real-world decision-support model and showing how a TOO approach can help identify latent structure in a stochastic simulator, as applied to outbreak surveillance. The \texttt{adaptive\_ts} package is under active development, with ample opportunities for expansion, such as handling multi-output simulators, competing objectives for optimization, and implementation of cutting-edge emulators to model and explore high-dimensional nonlinear dynamical systems.

\section*{ACKNOWLEDGMENTS}
This material is based upon work supported by the National Science Foundation under Grant 2200234, the National Institutes of Health under Grants R01AI158666 and R01DA055502, the U.S. Department of Energy, Office of Science, under contract number DE-AC02-06CH11357, and the Bio-preparedness Research Virtual Environment (BRaVE) initiative.

% Reducing font size (to 9pt) for References & Author Biagraphies
\footnotesize

% Please don't exchange the bibliographystyle style
\bibliographystyle{wsc}

% AUTHOR: Include your bib file here
\bibliography{refs}

\end{document}